\documentclass[granma]{svjour}

\usepackage[latin1]{inputenc}
\usepackage{graphicx}

\newcommand{\vt}{v_\mathrm{t}}

\newcommand{\Ftmax}{F_\mathrm{t,max}}

\begin{document}
\title{Influence of particle elasticity in shear testers}
\author{Dirk Kadau\inst{1}\mail{kadau@comphys.uni-duisburg.de},
  Dominik Schwesig\inst{1,2},  J\"org Theuerkauf\inst{2} and Dietrich  E.~Wolf\inst{1} 
  \thanks{ We thank J\'anos T\"or\"ok for stimulating discussions and 
visualization of the shear bands.
This work was supported by the German Science Foundation (DFG)
  within SFB 445 and grant
  WO 577/3-3,  and by Federal Mogul GmbH. } }
\institute{ ~\inst{1}Department of Physics \\
  Duisburg-Essen  University \\ D-47048~Duisburg, Germany \\ \\  ~\inst{2} The Dow Chemical Company \\
  Solids Processing Laboratory\\ Freeport TX 77541, USA  \\ \\ PACS number(s):
   45.70.-n, 62.20.-x, 83.10.Rs}
\date{Received: \today / Revised version: date}

\maketitle

\begin{abstract}
  
  Two dimensional simulations of non-cohesive granular matter in a
  biaxial shear tester are discussed. The effect of particle
  elasticity on the mechanical behavior is investigated using two
  complementary distinct element methods (DEM): Soft particle
  molecular dynamics simulations (Particle Flow Code, PFC) for elastic
  particles and contact dynamics simulations (CD) for the limit of
  perfectly rigid particles.  As soon as the system dilates to form
  shear bands, it relaxes the elastic strains so that one finds the
  same stresses for rigid respectively elastic particles in steady
  state flow.  The principal stresses in steady state flow are
  determined. They are proportional to each other, giving rise to an
  effective macroscopic friction coefficient which is about 10 \%
  smaller than the microscopic friction coefficient between the
  grains.

\end{abstract}

\section{Introduction}
In powder technology the material flow properties are usually determined
experimentally, e.g.\ using shear testers \cite{schwedes2003}. The 
measurements provide input for various phenomenological continuum models 
that have been proposed in order to calculate
dense granular flows. These models contain
parameters whose connection to the properties of the grains is not yet 
understood. It is the aim of distinct element simulation methods (DEM) 
to establish the connection between the grain scale and the macroscopic 
behavior directly \cite{cundall71,cundall79}.

The stress-strain behavior of a dense granular assembly consists 
of two parts: the rearrangement of the particles on the one hand, and 
their individual elastic or plastic deformation on the other. 
In this paper we address the question,
how strongly the flow properties are influenced by the grain deformations
compared to particle rearrangements.
Therefore, two different distinct element
methods are used: soft particle molecular dynamics modelling elastic
particles (used here: PFC) and contact dynamics (CD) to simulate perfectly 
rigid particles. By comparing the results of the
two methods the influence of particle elasticity can be separated
from the effect that particle rearrangements have on the macroscopic
stress-strain behavior found in e.g.\ the biaxial shear tester considered 
here.  

We simulate dense granular flow in a biaxial tester, 
which allows for larger displacements 
than the Jenike shear cell. 
The biaxial shear tester is a 
rectangular box \cite{nowak94,janssenCET}
in which the material is sheared under
constant strain rate in one direction and constant stress in a 
perpendicular  direction while the load plates in the third direction are
fixed. In this setup one will reach steady state flow (constant volume and
stress tensor), and the yield locus can be determined.

\section{Models}\label{sec:models}

Both models we use simulate the
trajectory of each individual particle by integrating Newton's
equations. They mainly differ in the way, how the  contact
forces between grains are determined. In soft particle molecular
dynamics (PFC) microscopic elastic deformations of each particle have to be 
taken into account: They determine the forces. By contrast,
in contact dynamics (CD) particles are considered as perfectly rigid, 
and forces are calculated from the volume exclusion constraint.

In order to make the comparison between the two methods as stringent
as possible, we simulate a very simple two dimensional system of
noncohesive round particles (disks) with Coulomb friction. The
boundary conditions and initial particle configurations are the same
for both types of simulations.  In both cases the normal restitution
coefficient is zero and the friction coefficient is $\mu=0.3$. In
order to give specific numbers we also introduce dimensional
parameters, which have no effect on the simulations: The average
particle radius is 1 mm, the mass density of the material the
particles are made of is $10^3$ kg/m$^3$. Whereas no further
parameters enter CD, the molecular dynamics algorithm requires the
specification of two stiffness parameters as the particles are not
perfectly rigid.

\subsection{Molecular Dynamics Simulation}\label{sec:DEM}

The Particle Flow Code used here is based on soft particle
molecular dynamics \cite{cundall79,cundall82}. Deformations of colliding
particles are represented by the overlap of idealized disks, i.e.
interaction forces between the particles are functions of the overlap
(= sum of the two disk radii - distance between the disk centers). 

\begin{figure}
   \centering
   \includegraphics[scale=0.4]{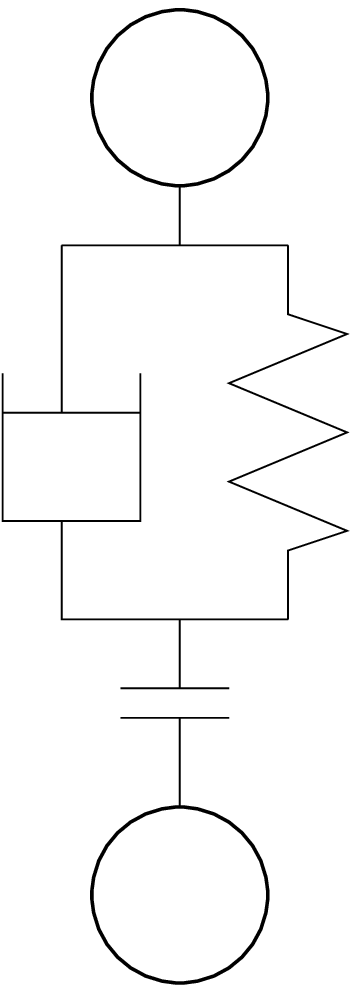}
   \includegraphics[scale=0.4]{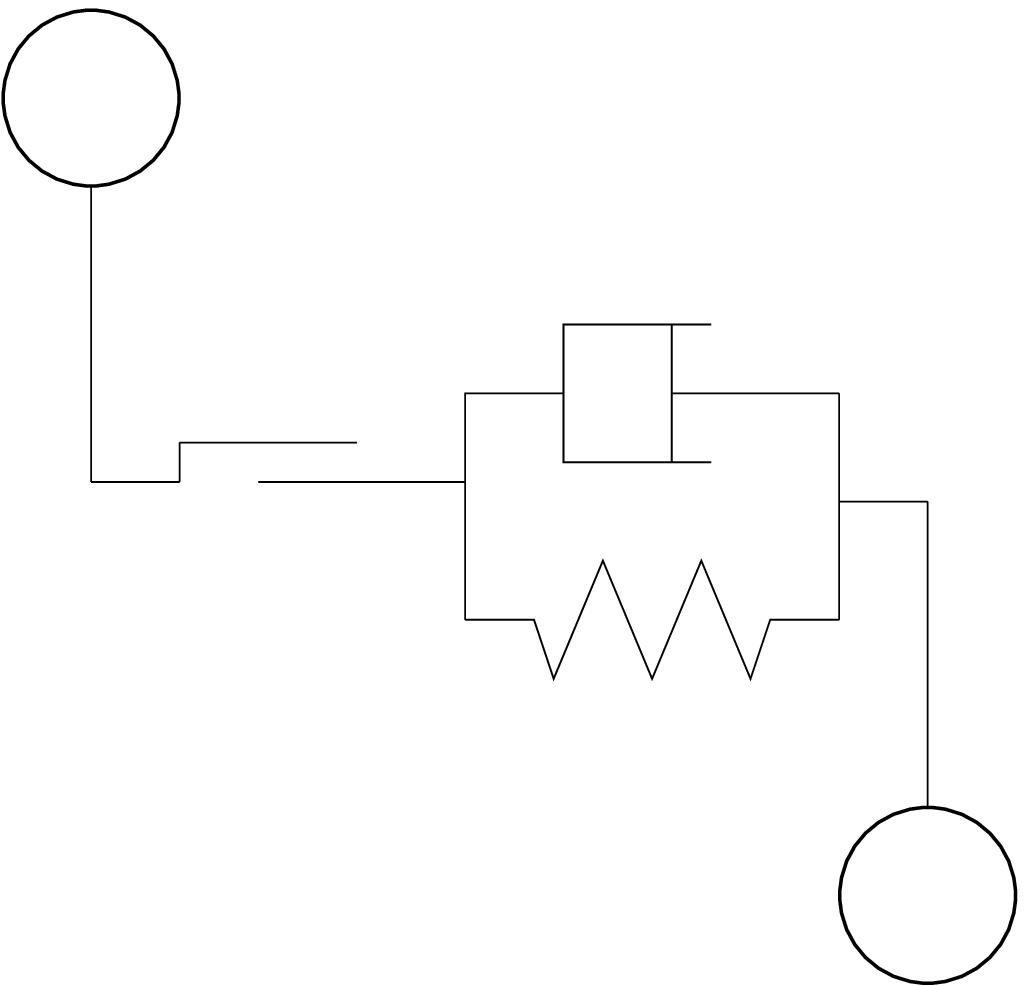}
   \caption{Mechanical contact model used in the PFC simulation: 
    Decomposition of contact forces
     between particles in normal (left figure) and tangential (right figure)
   direction. In both directions a spring and a dashpot is
     applied. Tensile forces  are prevented in normal direction (left
     figure),   in tangential direction a
   frictional slider is used for realizing  
   Coulomb friction     (right figure).  
}
   \label{fig:DEM_graph}
\end{figure}

The force between particles during contact is calculated with
mechanical elements such as springs and dashpots \cite{wolf96,luding2004}. 
In Fig.~\ref{fig:DEM_graph} a basic visco-elastic contact model for two 
particles in contact is depicted schematically. The
contact force is decomposed into a normal and a tangential component. 
In the simplest case used here the normal force is 
assumed to depend linearly on the overlap (the displacement of the spring).
The dashpot contributes a normal dissipative force proportional 
to the time derivative of the overlap. The sum of both contributions 
$F_{\rm n}$ 
is restricted to be repulsive, i.e. tensile normal forces are not allowed, 
because the particles are assumed to be non-cohesive. 

The tangential component of the contact force is implemented in terms of 
a linear spring-dashpot as well,
where the displacement of the spring is the integral of the tangential
relative velocity over the time of nonzero overlap. This represents static
friction, hence it is restricted to absolute values smaller than 
$\mu F_{\rm n}$. When this threshold is reached, the tangential relative 
motion is regarded as sliding with sliding friction $\mu F_{\rm n}$ (directed
opposite to the tangential relative velocity).

The boundary conditions of the system are realized by either strain driven
walls or stress driven walls without any friction.
 For all of the PFC simulations the commercial
Particle Flow Code (PFC2d ver3.0 \cite{PFC2dmanual}) was used.

For the PFC simulations, disks and walls with normal
and tangential stiffness coefficients of 1 N/m
are used. This low stiffness has been chosen 
in order to emphasize the possible difference
compared to the perfectly rigid particles in CD (stiffness $\to\infty$). 
Moreover this keeps the computing time for PFC low (it
increases with the square root of the stiffness). The viscous damping
coefficient is set equal to the critical damping factor at each contact
so that the restitution coefficients are zero for all particle contacts.

\subsection{Contact Dynamics}\label{sec:cd}

The contact dynamics simulation method has been used for the
simulation of dense and dry granular materials since the beginning of
the 1980's. Presuming that the behavior of these materials is governed
by perfect volume exclusion and static Coulomb
friction \cite{loetstedt82,jean92,jean99,moreau94,unger2002b,brendel2004},
CD implements them as constraints without modelling the microscopic
elastic grain deformations underlying them. 
Although CD can deal with more general systems as well
\cite{kadau2003} we restrict ourselves to the simplest case here.

\begin{figure}
   \centering
   \includegraphics[width=\columnwidth]{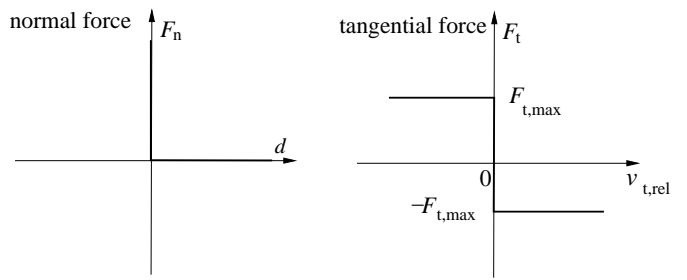}
   \caption{Contact laws in Contact Dynamics: Perfect volume exclusion
   (Signorini graph, left figure) determines the normal force $F_N$, the
   tangential force $F_t$ is determined by  exact
   implementation of Coulomb's law of friction with threshold $F_{\rm
   t,max}$ (right figure).  
}
   \label{fig:sig_coul_graph}
\end{figure}
Volume exclusion of perfectly rigid particles is characterized by the
Signorini graph, Fig.\ \ref{fig:sig_coul_graph}(left): When two
particles touch each other ($d=0$) the normal force $F_{\rm n}$ at the
contact assumes whatever positive value is needed to avoid
interpenetration.  This value can in principle be arbitrarily large
(Fig.\ \ref{fig:sig_coul_graph}, left). Otherwise the particles do not
interact. The tangential force $F_{\rm t}$ at a contact is determined by the
Coulomb friction law, Fig.\ \ref{fig:sig_coul_graph} (right): A
sticking contact (with relative tangential velocity $\vt=0$) can bear
any tangential force with absolute value up to a threshold $\Ftmax=\mu
F_{\rm n}$. Exceeding this threshold leads to sliding of the two particles
in contact. If sliding occurs ($\vt\neq 0$) the magnitude of the
tangential force is $\Ftmax$, while its direction is opposite to the
sliding velocity.

For dense systems consisting of many
particles with a complex contact network,  the calculation of all
contact forces is a global problem, because every force at a contact
influences all other contact forces.  There exists no analytical
solution for this complex problem \cite{radjai96,unger2004}, so that 
an iterative
procedure is applied: The forces at every contact are calculated in a
random order repeatedly until they all comply with each other.
In the simulations presented here a fixed number of iterations 
is chosen, which according to our previous experience \cite{unger2002} 
is large enough that the forces have converged within
a small tolerance. If the number of iterations would be too small, 
quasielastic effects would occur \cite{unger2002}.

\section{Simulated System} \label{sec:system}

\begin{figure}
\centering
 \parbox{0.1\columnwidth}{\includegraphics[width=0.1\columnwidth]{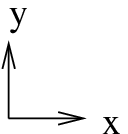} }
  \parbox{0.45\columnwidth}{ \includegraphics[width=0.45\columnwidth]{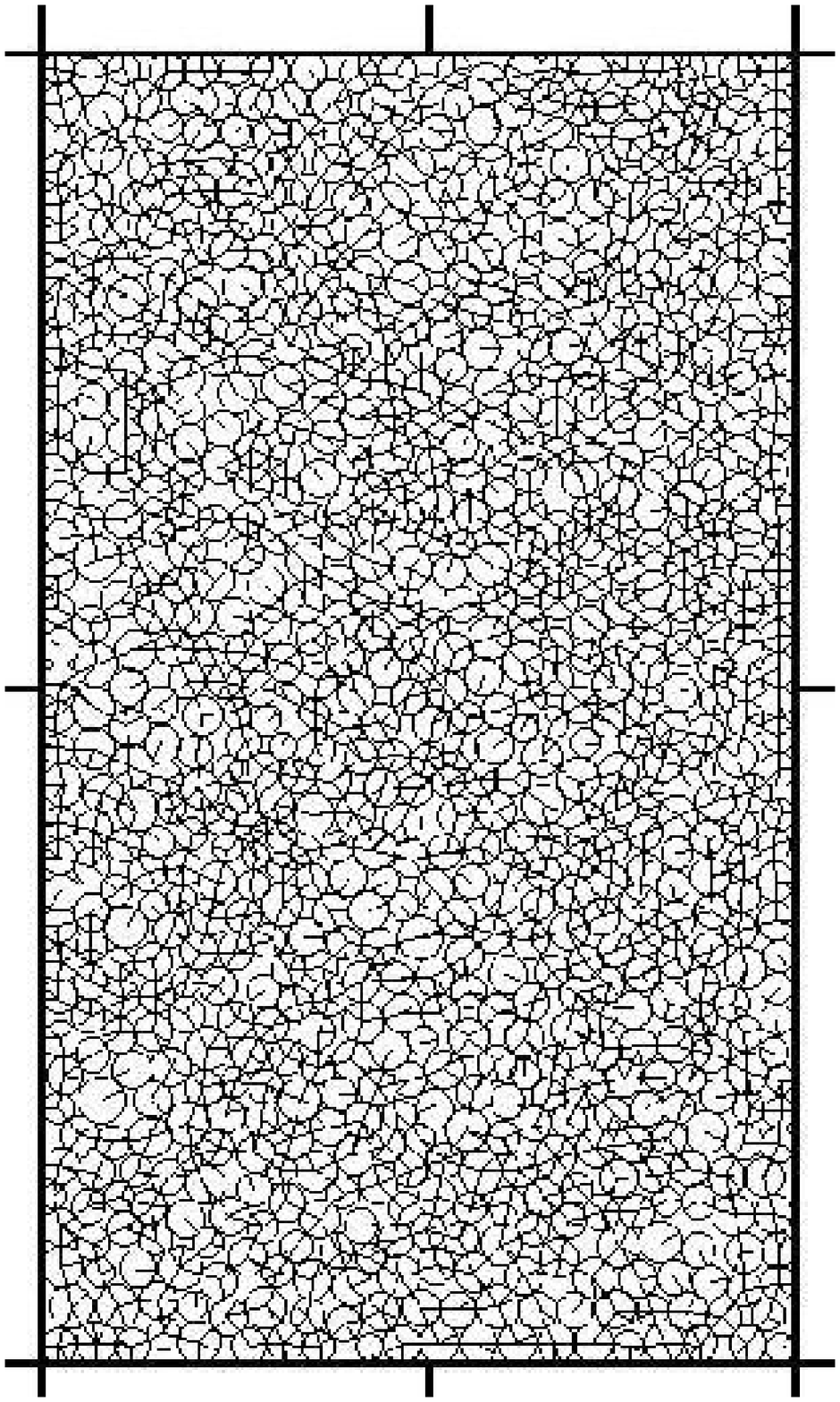}}
 \parbox{0.42\columnwidth}{ \includegraphics[width=0.419\columnwidth]{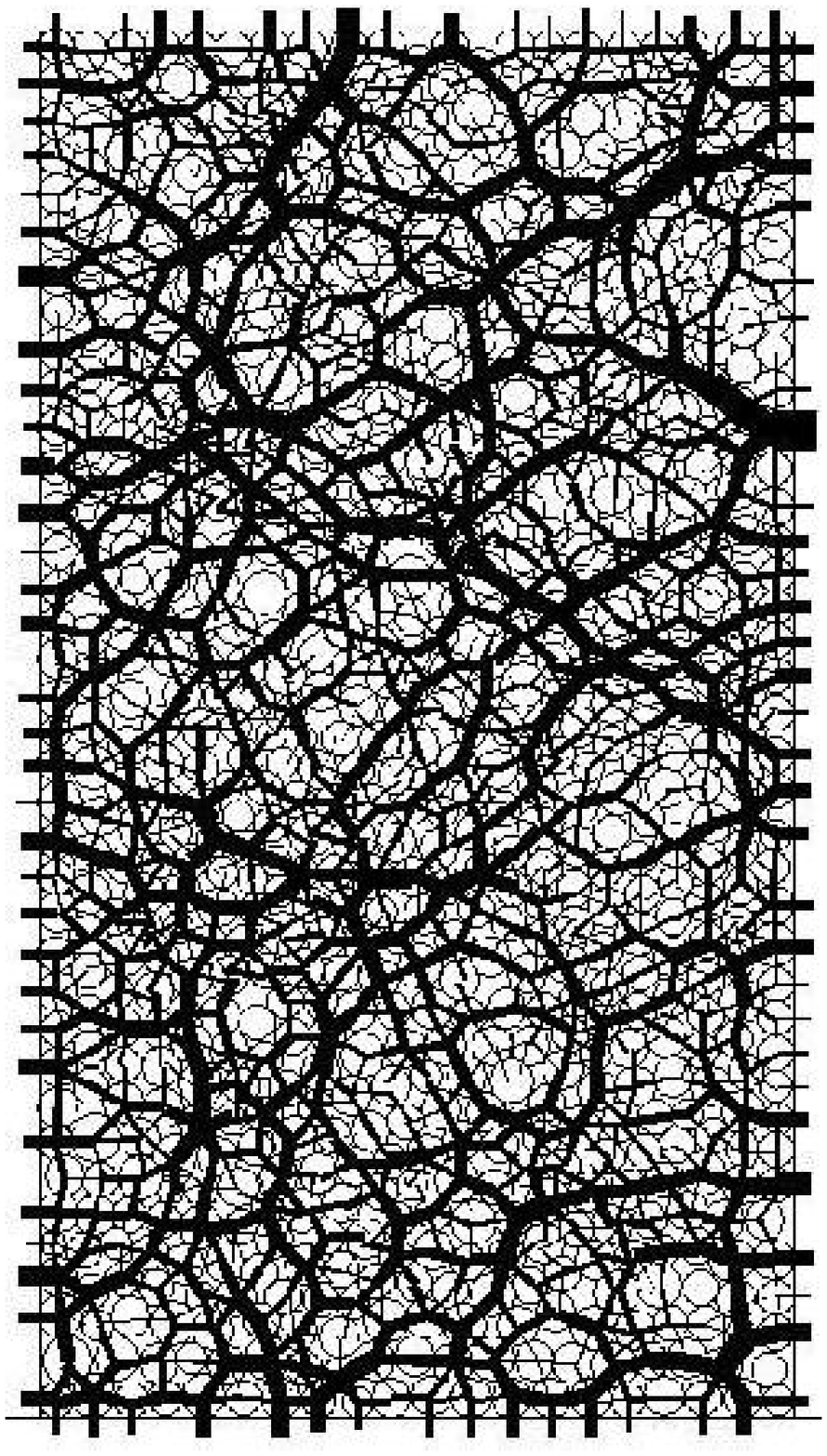}}
 \caption{Initial configurations used for CD (left) and
 PFC (right)  for shearing under fixed strain rate in $y$-direction
 and fixed stress in $x$-direction. In the right figure the left and
 right walls have been moved together leading to elastic response of
 the particles in order to fix the stress $\sigma_{xx}$. }
 \label{fig:Initial}
\end{figure}

We simulate a rectangular system confined by frictionless 
walls perpendicular to
the x- and y-axes, Fig.\ \ref{fig:Initial}. The boundary conditions
are such that the material is sheared under constant stress
$\sigma_{xx}$ at the yielding walls (perpendicular to the x-axis) and
with a constant strain rate $\dot\epsilon_{yy}$ at the pushing walls
(perpendicular to the y-axis). The stress at the pushing walls, $\sigma_{yy}$, 
and the strain rate at the yielding walls, $\dot\epsilon_{xx}$, are evaluated.
Since there is no wall friction in our simulations, 
$\sigma_{yy}$ and $\sigma_{xx}$ are the principal stresses. 

In the PFC and CD simulations the same initial configurations are
used. The system consists of about 800 round particles (disks) with a
Gaussian distribution of diameters cut off below $1/2$ and above $3/2$
of the average value. The width of the distribution is $0.1$ times the
average particle diameter.  The polydispersity is important to avoid
layering within the system.
 
 The initial configuration is prepared in the following way: The
 particles are randomly distributed at low concentration at first.
 Using contact dynamics the system is then biaxially compacted under
 small constant forces on the four walls (leading to stresses small
 compared to the ones applied later). This leads to a dense
 packing (Fig.\ \ref{fig:Initial}, left). The
 forces in $x$-direction were twice as large as those in
 $y$-direction in order to obtain an aspect ratio of about 2:1. 
If the average particle radius is taken as 1 mm, the system size is about 
43mm ($x$-direction) times 73mm ($y$-direction). Whereas
 in CD this configurations can be directly used as initial
 configuration ($t=0$), a further preprocessing step is needed for PFC.
Using molecular dynamics the configuration is compressed a little in
x-direction until the desired stress $\sigma_{xx}$
 is reached due to elastic response of the grains (here: small
 overlaps), while the extent in y-direction is kept constant.
Therefore the initial volume of the PFC-simulation (Fig.\ \ref{fig:Initial}, 
right) is about 6\% smaller in the x-direction than the one of the 
CD-simulation, but otherwise the initial configurations are the same.

 %

To reduce fluctuations  the results were averaged over 10 similar systems
where particles are randomly dispersed and located while
keeping the same average distribution.

\section{Results}

These dense systems (Fig.\ \ref{fig:Initial}) are
sheared under a constant strain rate $\dot\epsilon_{yy}=-0.02$ s$^{-1}$ and
different values of constant stress $\sigma_{xx}$.  The stresses are 
calculated by dividing the force on the corresponding wall by the wall area, 
where the average disk radius, 1 mm, is taken as size perpendicular to the 
xy-plane.

\begin{figure}
\centering
\includegraphics[width=0.9512195122\columnwidth]{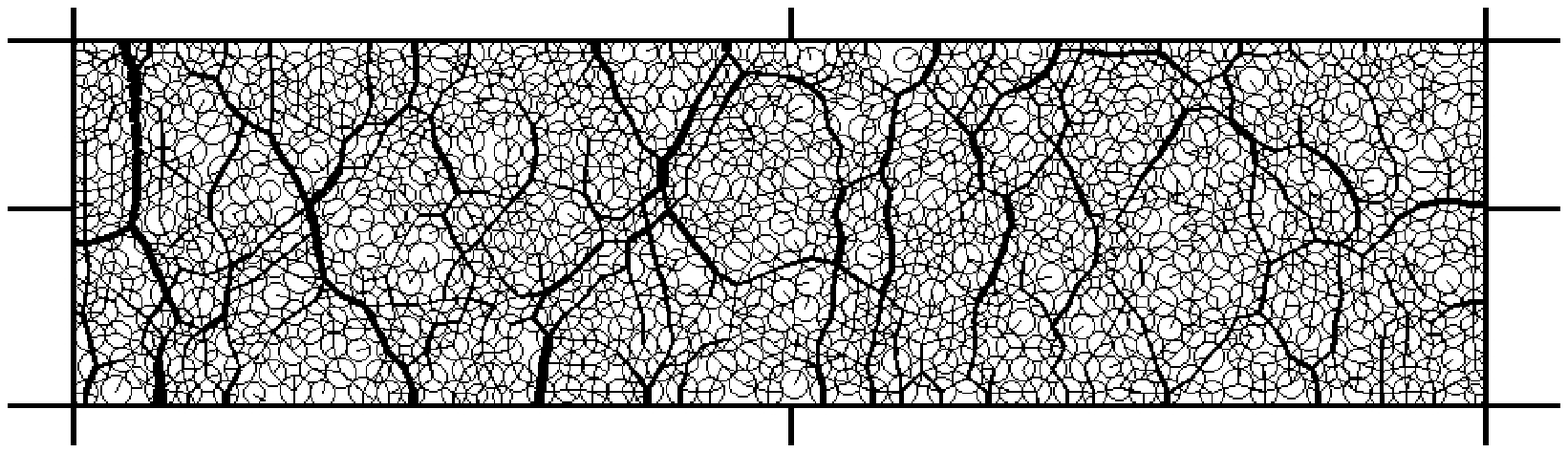}
\includegraphics[width=\columnwidth]{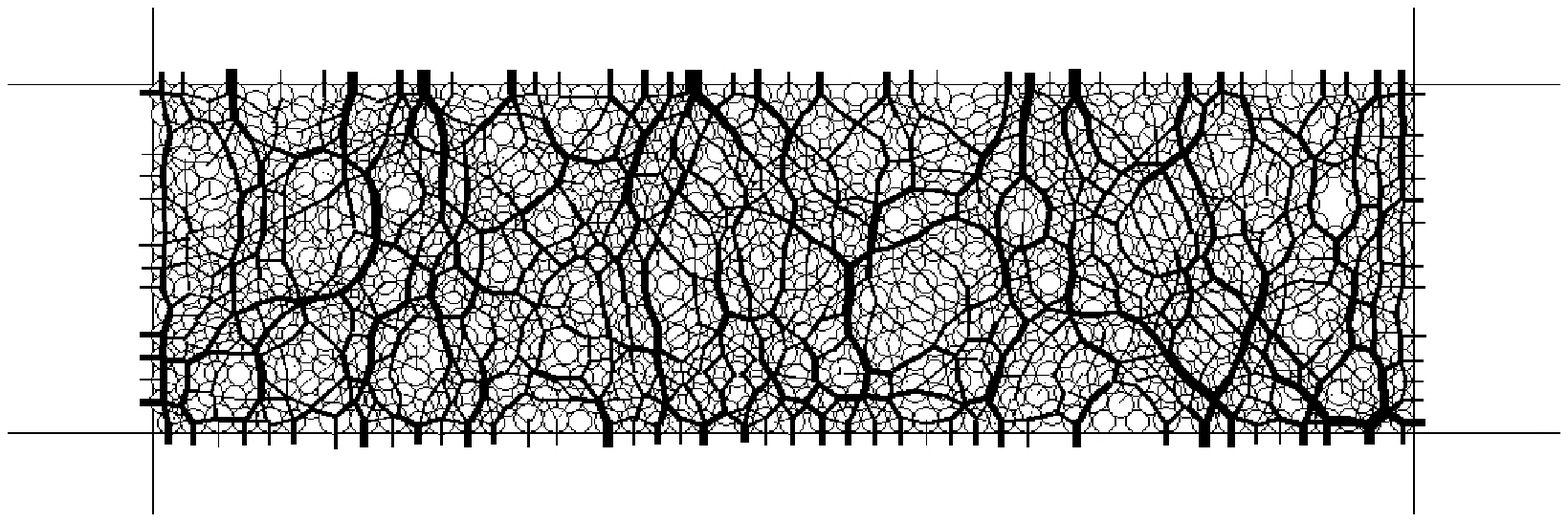}
 \caption{Typical configuration in steady state flow at the same time
   for CD (upper) and for PFC (lower panel). Both simulations started 
   with the same configuration (up to a horizontal compression by about 
   6\% for PFC, see text).} 
\label{fig:final}
\end{figure}

\begin{figure}
\centering
  \includegraphics[width=0.7\columnwidth]{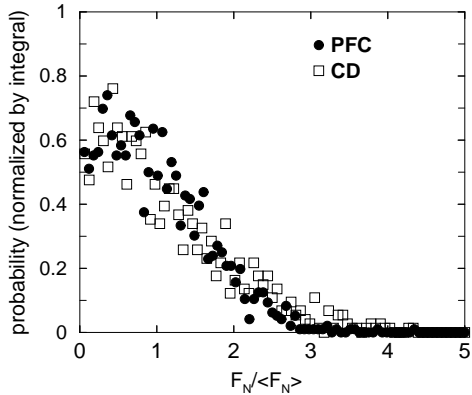}
\caption{Comparison of the distribution of normal forces from  a
  snapshot during steady state flow for CD and PFC.}
\label{fig:force_distribution}
\end{figure}

For $\sigma_{xx}=10$ Pa an example of a system after shearing for
$t=42$ s is shown in Fig.~\ref{fig:final}. As will be discussed below,
at this time steady state flow has already been established. The
vertical size of the simulation boxes is the same, because the strain
rate $\dot\epsilon_{yy}$ is the same in both simulations. Remarkably,
the horizontal size difference is still about 6\%, as for the initial
configurations, although it could have evolved differently depending
on the elastic properties, and in fact transiently does so. The strain
built up temporarily due to elasticity vanishes again in steady state
flow.  The final particle positions are different in the two
simulation methods, although they were initially the same, which is no
surprise in view of the highly nonlinear dynamics.  In particular the
force network evolves differently. On the other hand, Fig.
\ref{fig:force_distribution} compares the distributions of normal
forces for a snapshot like Fig.~\ref{fig:final}.  The
distributions obtained with CD and PFC are similar, thus the force
statistics in the systems are comparable.

We evaluated the velocities of the grains for the CD-simulations and
found shear localization in shear bands. In contrast to previous
simulations \cite{luding2004b,luding2005,luding2005b}, where the left wall was fixed, i.e.
strain controlled, whereas the right one was stress controlled, the
shear band which forms initially, is oriented along either diagonal,
breaking the symmetry of our system spontaneously.  Later on most of
the time four shearbands exist which form roughly a parallelogram
around the center which hardly moves. The dynamics of the shear bands
is very complex showing eddies and strong fluctuations. This will be
analyzed in more detail in a forthcoming publication \cite{torok2005}. 

\begin{figure}
\centering
  \includegraphics[width=0.45\columnwidth]{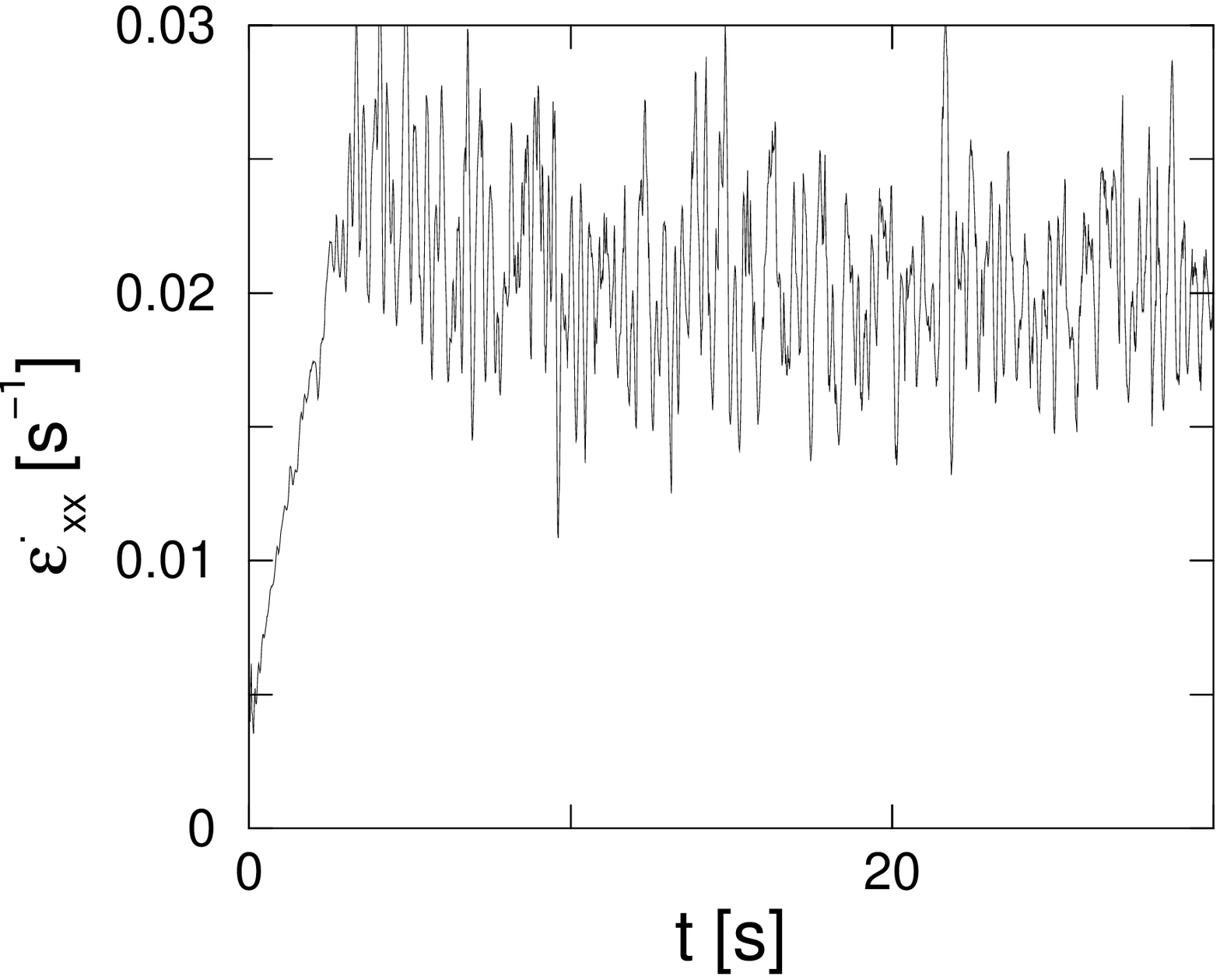}
  \includegraphics[width=0.45\columnwidth]{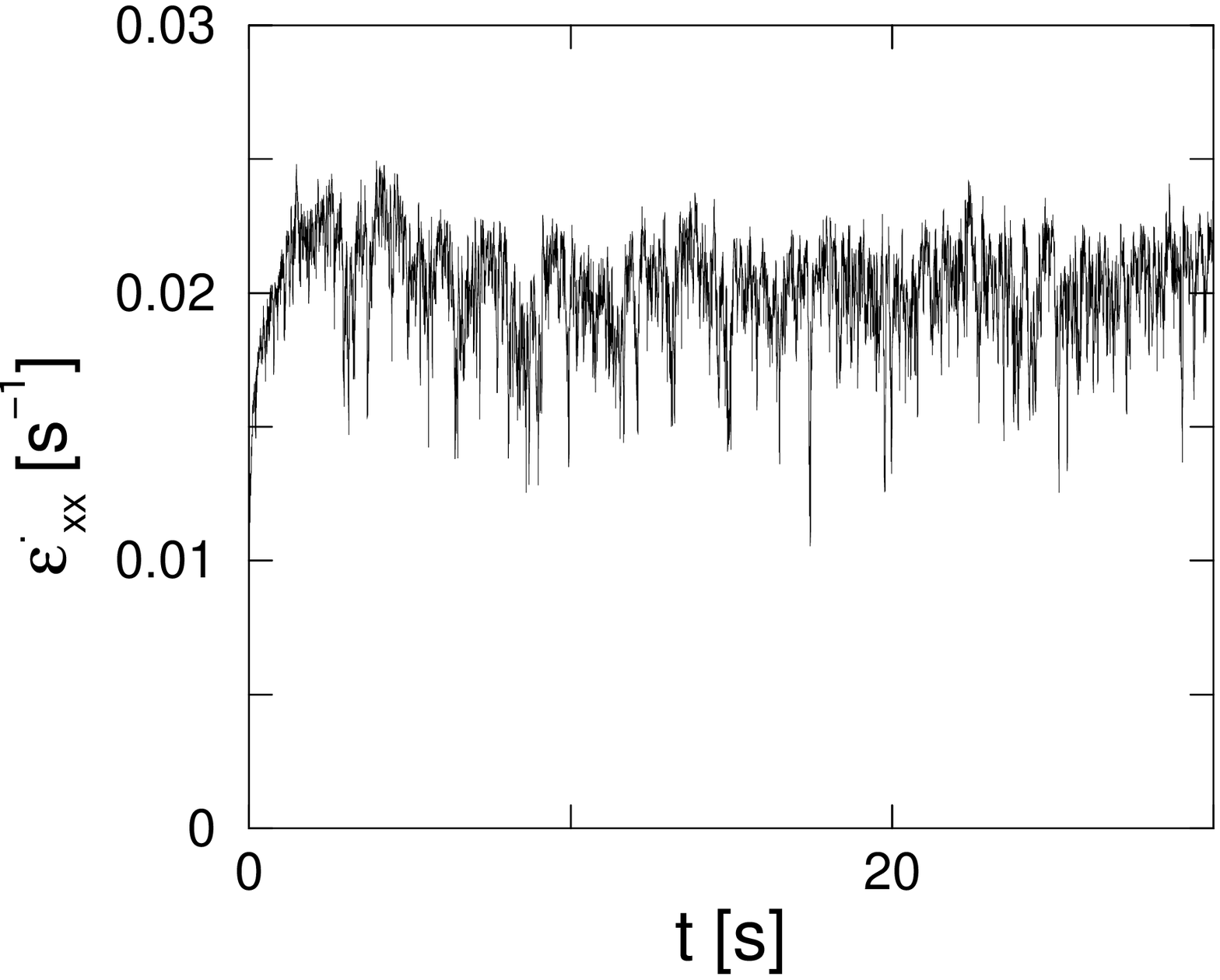}
 \caption{\label{fig:shear_strain}  Time dependence of the the strain rate
  $\dot\epsilon_{xx}$  at  fixed $\dot\epsilon_{yy}=-0.02$ s$^{-1}$ and
   $\sigma_{xx}=10$ Pa in comparison for PFC (left) and CD (right). 
   Strain curves for 
   different fixed  $\sigma_{xx}$ show the same behavior (not shown).
   } 
\end{figure}

The measured strain rate $\dot\epsilon_{xx}$, averaged over 10 similar
systems is shown in Fig.\ \ref{fig:shear_strain}. One can distinguish
three time intervals: During the first 2.5 s the elastic particles are
compressed, i.e.  $\dot\epsilon_{xx}<-\dot\epsilon_{yy}=0.02 {\rm
s}^{-1}$.  As expected, this region is much smaller for CD, where the
system of rigid disks gets compacted for only about 1 s, mostly due to
numerical errors creating tiny overlaps, which for perfect convergence
of the force iterations would not occur. However, a small amount of
compaction due to particle rearrangements is also possible.  The
second time interval extends up to 10 s in both simulation models and
is characterized by $\dot\epsilon_{xx}>-\dot\epsilon_{yy}$,
which means that the system dilates. The volume increases to
allow for shearing \cite{reynolds1885}. As already seen in Fig.~\ref{fig:final}
the elastic system dilates more, compensating for the initial compression, 
so that finally the volumes of both systems differ by roughly the same 
percentage as initially. Obviously the elastic energy stored during the
compression phase is completely released during dilation.   
After this transient the
system reaches a steady state where the strain rate fluctuates around
the average value $\langle \dot\epsilon_{xx}\rangle = -
\dot\epsilon_{yy}$. In this region the average volume remains
constant. The PFC-results agree qualitatively with the ones obtained with
molecular dynamics simulations by Luding
\cite{luding2004b,luding2005,luding2005b}. 

\begin{figure}
\centering
  \includegraphics[width=0.7\columnwidth]{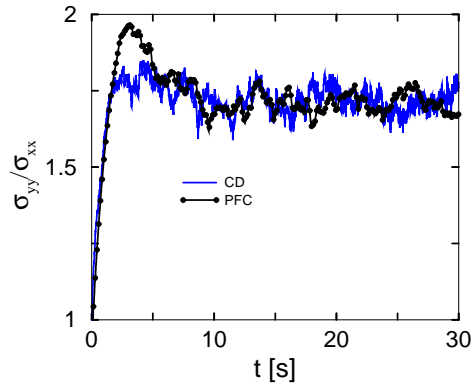}
 \caption{\label{fig:shear_stress} Time dependence of the the stress
  $\sigma_{yy}$  at  fixed $\dot\epsilon_{yy}=-0.02$ s$^{-1}$ and
   $\sigma_{xx}=10$ Pa for PFC and CD. }
\end{figure}

The time dependence of the stress $\sigma_{yy}$ is shown in 
Fig.~\ref{fig:shear_stress}: For elastic particles (PFC) it reaches
a pronounced maximum at the end of the compression phase after about 3~s.
During the dilatancy phase it decreases and finally fluctuates around a
constant value in steady state flow. For rigid particles (CD) the stress
maximum is not very pronounced: In order to distinguish it convincingly 
from the fluctuations, one would have to average over more than 10 independent
runs. This confirms that the stress maximum is due to elastic compression of
the system. On the other hand, the average steady state values 
of $\sigma_{yy}$ are the same for PFC and CD within the error bars.
Elastic contributions to $\sigma_{yy}$ cannot be observed any more
in agreement with our picture, that the elastic deformation relaxes during
dilation at the onset of shearing. 
    
\begin{figure}
\centering
  \includegraphics[width=0.7\columnwidth]{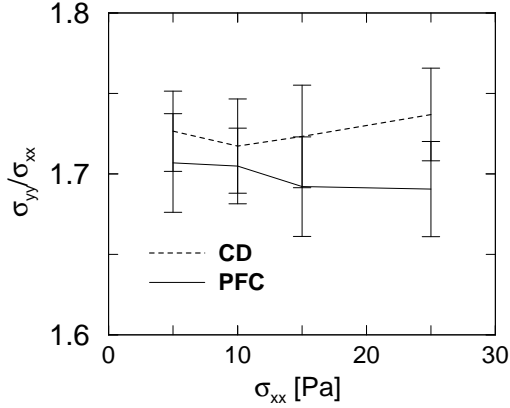}
 \caption{\label{fig:flowstresses} The average stress
  $\sigma_{yy}$ in steady state flow is proportional to 
  the applied stress $\sigma_{xx}$. 
}
\end{figure}

We repeated the simulations for four different stresses $\sigma_{xx}$
between 5 and 25 Pa keeping the strain rate $\dot\epsilon_{yy}=-0.02 
{\rm s}^{-1}$ the same.
The time-dependence of the  strain rate $\dot\epsilon_{xx}$ as well as
of the  stress ratio $\sigma_{yy}/\sigma_{xx}$ is in all cases
the same as in Fig.\ \ref{fig:shear_strain} respectively 
Fig.\ \ref{fig:shear_stress}. In particular,  
$\langle\sigma_{yy}\rangle/\sigma_{xx}$ time-averaged in steady state flow
does not depend on $\sigma_{xx}$, see Fig.\ \ref{fig:flowstresses}.
\footnote{Although the average values for CD seem to be slightly
larger than those for PFC by 1 to 3 \%, the error bars are so big that
the above statement that elasticity does not play a significant role
in steady state flow is still valid.}

As both principal values of the stress tensor are proportional to each
other in steady state flow, this implies a linear effective yield locus.
The effective friction angle $\phi$ is given by
\begin{equation}
\sin{\phi}=\frac{\langle \sigma_{yy} \rangle - \sigma_{xx}}
{\langle \sigma_{yy} \rangle + \sigma_{xx}}.
\end{equation} 
With $\langle \sigma_{yy} \rangle/\sigma_{xx}\approx 1.71 \pm 0.05$ this 
implies an effective friction coefficient 
\begin{equation}
\mu_{\rm eff} = \tan{\phi} \approx 0.27 \pm 0.02
\end{equation}
which is slightly smaller than the friction coefficient between the grains, 
$\mu = 0.3$.

The effect of system size was investigated by increasing the number of
particles to 10000. However, no effect of the system size on the
stress ratio was found.

\section{Conclusion}

The results presented above for a two dimensional model of a biaxial
shear test show that particle elasticity only affects
the stress and volumetric strain during a short transient. Dilatancy
leads to elastic strain relaxation so that the stress in steady state flow is 
essentially determined by rearrangements of the particles.
This could be proven by
comparing the simulation results obtained for elastic and perfectly
rigid particles, respectively.   

In steady state flow the principal stresses turned out to be
proportional to each other. Using Mohr-Coulomb theory
we determined the effective macroscopic friction coefficient
of the granular material, which is the ratio of the shear stress to the
normal stress at a shear plane. We could relate it to the microscopic
friction coefficient between the grains, which is slightly larger.

We  also checked the influence of the strain rate $\dot\epsilon_{yy}$
and found essentially no effect. This can be explained by comparing
the orders of magnitude of inertia forces due to the prescribed strain rate
to the
forces in the system caused by the applied stress $\sigma_{xx}$. The
typical time scale for inertia effects is given by the inverse strain
rate $1/\dot\epsilon_{yy}$, the length scale is given
by the system size $L_y$, the typical mass is the average particle
mass $\rho r_0^3$ (we neglect the factor $\pi$ here). Thus, typical
inertia forces are: $F_{\rm inertia}\sim\rho r_0^3 L_y
\dot\epsilon_{yy}^2$. Typical forces  due to the applied stress
$\sigma_{xx}$ are estimated by multiplying with the system size $L_y$
(and its thickness $r_0$):
$F_{\rm ext} \sim \sigma_{xx} L_y r_0$. The interesting ratio: $F_{\rm
  inertia}/F_{\rm ext}\sim \rho r_0^2\dot\epsilon_{yy}^2/\sigma_{xx}$. 
Inserting the used values results in: $F_{\rm
  inertia}/F_{\rm ext} \sim 4\cdot 10^{-8}$ which is a small number
so that we are well in the region of slow (quasistatic) 
deformations where we do not
expect an influence of the applied strain rate on the results as is
found here.
This argument shows, that only the dimensionless ratio 
$\rho r_0^2\dot\epsilon_{yy}^2/\sigma_{xx}$ matters, as long as elastic
effects are negligible: Increasing $\dot\epsilon_{yy}^2$ is equivalent 
to decreasing $\sigma_{xx}$.

\bibliographystyle{unsrt}
\bibliography{paper_elasticity}

\begin{thebibliography}{10}

\bibitem{schwedes2003}
J.~Schwedes.
\newblock Review on testers for measuring of properties of bulk solids.
\newblock {\em Granular Matter}, 5:1--43, 2003.

\bibitem{cundall71}
P.A. Cundall.
\newblock A computer model for simulating progressive large scale movements of
  block rock systems.
\newblock In {\em Proceedings of the Symposium of the International Society of
  Rock Mechanics}, volume~1, pages 132--150, Nancy, France, 1971.

\bibitem{cundall79}
P.A. Cundall and O.D.L Strack.
\newblock A discrete model for granular assemblies.
\newblock {\em Geotechnique}, 29:47--65, 1979.

\bibitem{nowak94}
M.~Nowak.
\newblock {\em Spannungs-/Dehnungsverhalten von Kalkstein in der Zweiaxialbox}.
\newblock PhD thesis, TU Braunschweig, 1994.

\bibitem{janssenCET}
R.~J.~M. Janssen and H.~Zetzener.
\newblock Measurements on cohesive powder with two biaxial shear testers.
\newblock {\em Chemical Engineering \& Technology}, 26(2):147--151, ??

\bibitem{cundall82}
P.A. Cundall.
\newblock Distinct element models of rock and soil structure.
\newblock In {\em Analytical and computational methods in engineering and rock
  mechanics}, London, England, 1987. Allen \& Unwin.

\bibitem{wolf96}
D.E. Wolf.
\newblock Modelling and computer simulation of granular media.
\newblock In K.H. Hoffmann and M.~Schreiber, editors, {\em Computational
  Physics: Selected Methods - Simple Exercises - Serious Applications}, pages
  64--94, Heidelberg, 1996. Springer.

\bibitem{luding2004}
S.~Luding.
\newblock Molecular dynamics simulations of granular materials.
\newblock In H.~Hinrichsen and D.~E. Wolf, editors, {\em The Physics of
  Granular Media}, Berlin, Germany, 2004. Wiley-VCH.

\bibitem{PFC2dmanual}
Itasca consulting, www.itasca.com.
\newblock {\em Particle Flow Code in 2 Dimensions}.
\newblock Online manual PFC2d version 3.0.

\bibitem{loetstedt82}
P.~Lötstedt.
\newblock Mechanical systems of rigid bodies subject to unilateral constraints.
\newblock {\em SIAM J. Appl. Math.}, 42:281--296, 1982.

\bibitem{jean92}
M.~Jean and J.~J. Moreau.
\newblock Unilaterality and dry friction in the dynamics of rigid body
  collections.
\newblock In {\em Proceedings of Contact Mechanics International Symposium},
  pages 31--48, Lausanne, Switzerland, 1992. Presses Polytechniques et
  Universitaires Romandes.

\bibitem{jean99}
M.~Jean.
\newblock The non-smooth contact dynamics method.
\newblock {\em Comput. Methods Appl. Engrg.}, 177:235--257, 1999.

\bibitem{moreau94}
J.J. Moreau.
\newblock Some numerical methods in multibody dynamics: application to granular
  materials.
\newblock {\em Eur J Mech, A/Solids}, 13(4):93--114, 1994.

\bibitem{unger2002b}
T.~Unger and J.~Kert\'esz.
\newblock The contact dynamics method for granular media.
\newblock In {\em Modeling of Complex Systems}, pages 116--138, Melville, New
  York, 2003. American Institute of Physics.
\newblock cond-mat/0211696.

\bibitem{brendel2004}
L.~Brendel, T.~Unger, and D.~E. Wolf.
\newblock Contact dynamics for beginners.
\newblock In H.~Hinrichsen and D.~E. Wolf, editors, {\em The Physics of
  Granular Media}, Berlin, Germany, 2004. Wiley-VCH.

\bibitem{kadau2003}
D.~Kadau, G.~Bartels, L.~Brendel, and D.~E. Wolf.
\newblock Pore stabilization in cohesive granular systems.
\newblock {\em Phase Trans.}, 76(4-5):315--331, 2003.

\bibitem{radjai96}
F.~Radjai, L.~Brendel, and S.~Roux.
\newblock Nonsmoothness, indeterminacy, and friction in two dimensional arrays
  of rigid particles.
\newblock {\em Phys. Rev. E}, 54(1):861, 1996.

\bibitem{unger2004}
T.~Unger, J.~Kert\'esz, and D.~E. Wolf.
\newblock Force indeterminacy in the jammed state of hard disks.
\newblock {\em Phys. Rev. Lett.}, 94:178001, 2005.

\bibitem{unger2002}
T.~Unger, L.~Brendel, D.~E. Wolf, and J.~Kert{\'e}sz.
\newblock Elastic behavior in contact dynamics of rigid particles.
\newblock {\em Phys. Rev. E}, 65(6):061305, 2002.

\bibitem{luding2004b}
S.~Luding, R.~Tykhoniuk, J.~Tomas, L.~Heim, M.~Kappl, and H.-J. Butt.
\newblock Flow behavior of cohesive and frictional fine powders.
\newblock In Y.~Shimizu, R.~D. Hart, and P.~A. Cundall, editors, {\em Numerical
  Modeling in Micromechanics via Particle Methods - 2004}, pages 157--163. A.
  A. Balkema, 2004.
\newblock PFC Symposium procedings.

\bibitem{luding2005}
S.~Luding.
\newblock Anisotropy in cohesive, frictional granular media.
\newblock {\em J. Phys.: Condens. Matter}, 17:S2623--S2640, 2005.

\bibitem{luding2005b}
S.~Luding.
\newblock Shear flow modeling of cohesive and frictional fine powder.
\newblock {\em Powder Technology}, 2005.
\newblock In press, corrected proof available online 23 May 2005.

\bibitem{torok2005}
J.~T\"or\"ok, D.~Kadau, and D.~E. Wolf.
\newblock Properties of shear bands in biaxial tests.
\newblock in preparation.

\bibitem{reynolds1885}
O.~Reynolds.
\newblock On the dilatancy of media composed of rigid particles in contact.
\newblock {\em Philos. Mag.}, Ser. 5(20):469, 1885.

\end{thebibliography}

\end{document}